\documentclass{mem}
\usepackage{natbib}\usepackage{txfonts}\usepackage{balance}
\usepackage{graphicx}
\usepackage[a4paper]{hyperref}
\idline{75}{282}
\begin{document}
\def\teff{$T\rm_{eff }$}
\def\kms{$\mathrm {km s}^{-1}$}

\title{
Primordial gas heating by dark matter and structure~formation
}

   \subtitle{}

\author{
M. \,Mapelli\inst{1} 
\and E. \, Ripamonti\inst{2}
          }

  \offprints{M. Mapelli}

\institute{
Institute for Theoretical Physics --
University of Zurich, Winterthurerstrasse 190,
CH-8057 Zurich, Switzerland
\email{mapelli@physik.unizh.ch}
\and
Kapteyn Astronomical Institute, University of Groningen, Postbus 800, 9700 AV, Groningen, The Netherlands
}

\authorrunning{Mapelli \and Ripamonti}

\titlerunning{Primordial gas heating by dark matter and structure formation}

\abstract{
Dark matter (DM) decays and annihilations might heat and partially reionize the Universe at high redshift. Although this effect is not important for the cosmic reionization, the gas heating due to DM particles might affect the structure formation. In particular, the critical halo mass for collapse is increased up to a factor of $\sim{}2$. Also the fraction of gas which collapses inside the smallest halos is substantially reduced with respect to the cosmological value. These effects imply that DM decays and annihilations might delay the formation of the first structures and reduce the total star mass in the smallest halos.
 
}
\maketitle{}

\section{Introduction}

Dark matter (DM) particles can either decay or annihilate, producing photons, neutrinos, electron-positron pairs, and/or more massive particles, depending on the mass of the progenitor.
In this way, DM particles indirectly interact with the intergalactic medium (IGM), transferring to it a part of their mass-energy. Then, DM decays and annihilations can influence the thermal and ionization history of the Universe.

The fraction of  energy produced by decays/annihilations which is effectively absorbed by the IGM ($f_{abs}(z)$) is a crucial ingredient in this scenario, and depends on the mass of the progenitor and on the nature of the product particles. If the product particles are  quite massive ($\gtrsim{}135$ MeV), they can generate a cascade, so that  predicting the behaviour of $f_{abs}(z)$ is difficult. However, \cite{map1} proved that ordinary cold DM (i.e. particles with mass greater than a few GeV) has negligible effects on the history of reionization and heating even assuming $f_{abs}(z)=1$.

Instead, in the case of lighter DM particles (such as sterile neutrinos and light dark matter) the effects on reionization might be important, if we assume $f_{abs}(z)=1$ \citep{map1}. Then, for lighter DM particles it is crucial to correctly estimate the behaviour of $f_{abs}(z)$.

Luckily, $f_{abs}(z)$ can be calculated (Ripamonti et al. 2007a; Mapelli \& Ripamonti 2007) if the product particles are photons, electron-positron pairs or neutrinos (which have negligible interactions). In fact, photons interact with the IGM via Compton scattering and photo-ionization; while pairs are involved in inverse Compton scattering, collisional ionizations and positron annihilations. The resulting $f_{abs}(z)$, both in case of photons and pairs, is close to 1 at high redshift ($z>>100$), whereas it drops to values $\lesssim{}0.1$ at low redshift. This result implies that DM is not an efficient source of cosmic reionization at redshift $\lesssim{}20$. However, DM decays and annihilations can heat and partially reionize the Universe already at very high redshift ($z\approx{}50-100$), when the first stars and structures have not appeared yet. 
In this proceeding, we will show that the heating and partial ionization produced at high redshift by DM decays and annihilations have an impact on the process of structure formation itself.

\section{Heating and molecular abundance enhancement}
In the following, we will consider three different scenarios: decays of sterile neutrinos with mass $4-25$ keV (top panel) into photons and active neutrinos, decays of light DM (LDM) particles with mass $3-10$ MeV (central) and annihilations of LDM with mass $1-10$ MeV (bottom) into pairs. Further details on these DM models can be found in \cite{ripa1}.

\begin{figure}[h!]
\resizebox{\hsize}{!}{\includegraphics[clip=true]{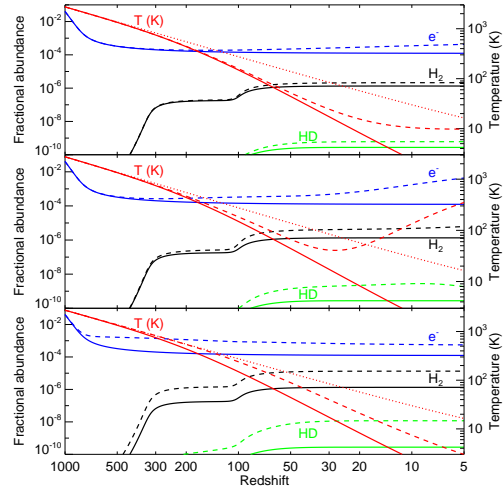}}
\caption{\footnotesize
Effects of decaying/annihilating DM on the IGM evolution. Left axis: fractional abundances of free electrons ($e^-$), H$_2$ and HD as a function of  redshift. Right axis: matter temperature as a function of  redshift. Top panel: Effect of decaying sterile neutrinos of mass 25 keV (dashed line).  Central panel:  decaying LDM of mass 10 MeV (dashed line).  Bottom panel: annihilating LDM of mass 1 MeV (dashed line). In all the panels the dotted line is the cosmic microwave background (CMB) temperature and the solid line represents the thermal and chemical evolution without DM decays/annihilations.
}
\label{fig:fig1}
\end{figure}

DM decays and annihilations increase both the temperature of the IGM and the fraction of free electrons (Fig.~\ref{fig:fig1}). Free electrons are also catalysts for the formation of molecular hydrogen (H$_2$) and HD molecules, whose fractional abundance is thus enhanced (Fig.~\ref{fig:fig1}).

H$_2$ and HD play a fundamental role in the primordial Universe, as they are the most efficient coolants in absence of metals. Then, if their abundance is increased, structure formation is favoured. On the other hand, the temperature increase due to DM decays/annihilations tends to delay structure formation.
 
\begin{figure}[h!]
\resizebox{\hsize}{!}{\includegraphics[clip=true]{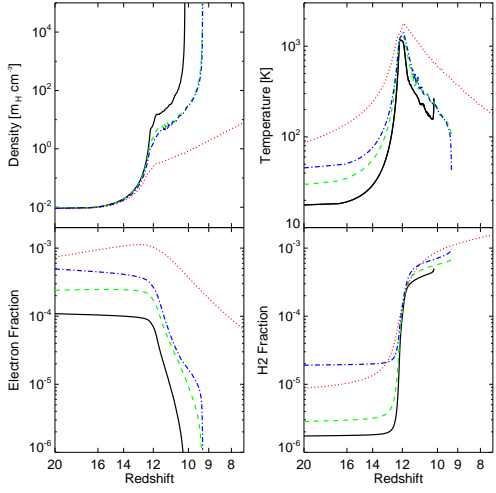}}
\caption{\footnotesize
Evolution of the central region of a
$6\times{}10^5\,{}M_\odot{}$ isothermal halo virializing at
$z_{vir}=12$.  From left to right and from top to bottom: density,
temperature, electron abundance and H$_2$ abundance as function of
redshift. The solid line represents the unperturbed case (i.e. without
DM decays and annihilations). The dashed, dot-dashed and dotted lines
account for the contribution of 25-keV sterile neutrino decays, 1-MeV
LDM annihilations and 10-MeV LDM decays, respectively. 
}
\label{fig:fig2}
\end{figure}

To understand which of these two effects dominates, we cannot limit our analysis to the IGM, where the density is uniform, but we have to study the chemical and dynamical evolution inside the first halos. To this purpose, we used a one-dimensional Lagrangian code \citep{ripa2}, which simulates the gravitational and hydro-dynamical evolution of the gas, accounting for the behaviour of 12 chemical species, for the cooling/heating effects and for the gravitational influence of DM halo. Fig.~\ref{fig:fig2} shows the evolution of gas density, temperature, ionization fraction and H$_2$ fractional abundances within a simulated halo of $6\times{}10^5\,{}M_\odot{}$ virializing at $z=12$. Black solid line is the case without DM decays/annihilations, while the colour lines account for DM decays/annihilations. From the density plot it is evident that  DM decays/annihilations delay the collapse, especially in the case of LDM annihilations (red dotted line). Thus, the temperature increase dominates over the coolant enhancement.

\section{The critical mass}
To quantify the importance of DM decays and annihilations in delaying the collapse we calculated the critical mass ($m_{crit}$), i.e. the minimum halo mass for collapse at a given redshift \citep{tegmark}.
\begin{figure}[h!]
\resizebox{\hsize}{!}{\includegraphics[clip=true]{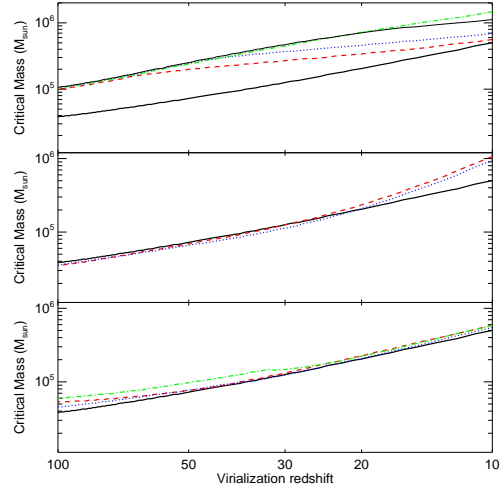}}
\caption{\footnotesize
 The critical mass as a function of the
virialization redshift $z_{vir}$ for isothermal DM halos. Top panel: Effect of decaying sterile neutrinos of mass 25 (dashed line), 15 (dotted) and 4 keV (dot-dashed). Central panel: Decaying LDM of mass 10 (dashed line) and 3 MeV (dotted). Bottom panel: Annihilating LDM of mass 10 (dashed line), 3 (dotted) and 1 MeV (dot-dashed). In all
the panels the solid line represents the thermal and chemical evolution without
DM decays/annihilations.  The thin solid line used in the top panel represents the
case of non-decaying warm DM (with mass 4 keV).
}
\label{fig:fig3}
\end{figure}
Fig.~\ref{fig:fig3} shows the behaviour of $m_{crit}$ in the case we assume an isothermal DM halo. The results do not change significantly for a \cite{NFW} profile. In presence of DM decays and annihilations $m_{crit}$ is always increased (Ripamonti et al. 2007b; Ripamonti \& Mapelli 2007). However, the increase of $m_{crit}$ in the case of sterile neutrinos is due not only to decays, but mainly to the fact that these particles are warm DM. In fact, given their large dispersion velocity, they prevent the formation of small and highly concentrated halos.

In the case of LDM decays and annihilations the increase of $m_{crit}$ is always less than a factor of $\sim{}2$. In the case of LDM decays the effect is maximum at relatively low redshifts ($z\lesssim{}20$, whereas in the case of LDM annihilations it is stronger at high redshift ($z>30$). This difference arises from the fact that the decay and the annihilation rate are proportional to the DM density and to the square of the DM density, respectively.

\section{The dearth of baryons in primordial halos}
In conclusion, DM decays and annihilations tend to increase $m_{crit}$; but this is a tiny effect, probably impossible to observe. However, DM decays and annihilations can have also an other effect on structure formation, possibly more important.

\begin{figure}[h!]
\resizebox{\hsize}{!}{\includegraphics[clip=true]{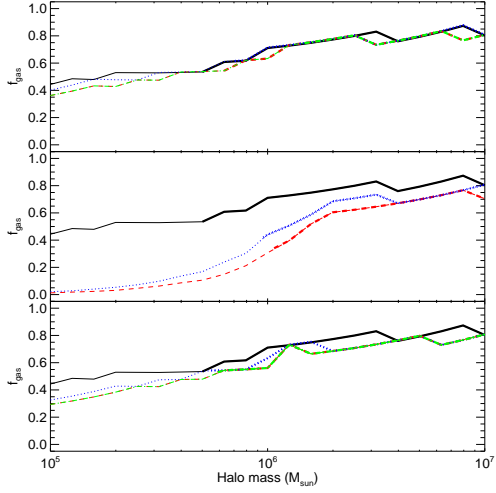}}
\caption{\footnotesize
 Halo baryonic mass fraction (see definition in
equation~\ref{eq:fraction_gas}) as a function of the halo mass for a
fixed virialization redshift ($z_{vir}=10$) and for an isothermal
DM profile. Top panel: Effect of decaying sterile neutrinos. Central:
Decaying LDM. Bottom: Annihilating LDM. The lines used in the three
panels are the same as in Fig.~\ref{fig:fig3}. Thick (thin) lines indicate that the halo mass is larger (smaller) than $m_{crit}$.
}
\label{fig:fig4}
\end{figure}
The temperature increase due to DM decays and annihilations leads also to an increase of the gas pressure, which counterbalances the gravitational potential of DM. Then, it is more difficult for the gas to sink at the center of the DM halo.

To quantify this effect, in Fig.~\ref{fig:fig4} we show the quantity $f_{gas}$, defined as 
\begin{equation}\label{eq:fraction_gas}
f_{gas}=\frac{M_{gas}(R_{vir})}{M_{{\rm
halo}}}\,{}\frac{\Omega{}_{M}}{\Omega{}_b},
\end{equation}
where $M_{gas}(R_{vir})$ is the mass of gas within the virialization radius ($R_{vir}$) of a halo of total mass $M_{{\rm halo}}$ and $\Omega{}_b/\Omega{}_{M}$ is the cosmological ratio of baryonic versus total matter in the Universe ($\Omega{}_b=0.042$, $\Omega{}_{M}=0.24$, Spergel et al. 2007). Then, $f_{gas}$ measures what is the real ratio (from our simulations) between baryonic and total mass in the halo, normalized to the cosmological value.

The more interesting case in Fig.~\ref{fig:fig4} are LDM decays, where $f_{gas}$ for halos with $M_{{\rm halo}}\sim{}m_{crit}$ is of the order of 0.3-0.4. This means that, in presence of DM decays, the real fraction of gas mass inside the virialized halo is much lower than the cosmological value.

Then, even if halos are still allowed to collapse, the quantity of gas at their center is a factor of $\sim{}2$ smaller than it is expected from cosmology. As the reservoir of gas is smaller, also the total mass of stars which  form within these low mass halos might be less than expected.
Further investigations might clarify whether this effect is significant for the formation of first stars.

\begin{acknowledgements}
MM is grateful to the Societ\`a Astronomica Italiana (SAIt) for awarding her the Tacchini Prize 2007. MM also thanks the DAVID group for the useful and pleasant collaboration.
\end{acknowledgements}

\bibliographystyle{aa}

\begin{thebibliography}{}

\bibitem[Mapelli et al. (2006)]{map1} Mapelli, M., Ferrara A.~\& Pierpaoli E., 2006, \mnras, 369, 1719

\bibitem[Mapelli \& Ripamonti (2007)]{map2} Mapelli, M.~\& Ripamonti, E. 2007, to appear in the proceedings of the 11th Marcel Grossmann Meeting held in Berlin, Germany, July 23-29 2006, arXiv:astro-ph/0701672v1

\bibitem[Navarro, Frenk \& White (1996)]{NFW} Navarro J. F., Frenk C. S., White S. D. M., 1996, \apj, 462, 563

\bibitem[Ripamonti (2007)]{ripa2} Ripamonti, E., 2007, \mnras, 376, 709

\bibitem[Ripamonti \& Mapelli (2007)]{ripmap} Ripamonti, E.~\& Mapelli, M. 2007, to appear in the proceedings of the 11th Marcel Grossmann Meeting held in Berlin, Germany, July 23-29 2006, arXiv:astro-ph/0701673v1

\bibitem[Ripamonti et al. (2007a)]{ripa1} Ripamonti, E., Mapelli, M.~\& Ferrara, A. 2007a, \mnras, 374, 1067

\bibitem[Ripamonti et al. (2007b)]{ripa3} Ripamonti, E., Mapelli, M.~\& Ferrara, A. 2007b, \mnras, 375, 1399

\bibitem[Spergel et al. (2007)]{spergel} Spergel, D. N. et al. 2007, \apj, accepted, arXiv:astro-ph/0603449v2

\bibitem[Tegmark et al. (1997)]{tegmark} Tegmark M., Silk J., Rees M. J., Blanchard A., Abel T., Palla F. 1997, \apj, 474, 1

\end{thebibliography}

\end{document}